\documentclass[aip,jcp,reprint,amsmath,amssymb]{revtex4-1}
\usepackage{graphicx}
\usepackage{dcolumn}
\usepackage{bm}
\usepackage[utf8]{inputenc}
\usepackage[T1]{fontenc}
\usepackage{mathptmx}
\usepackage{etoolbox,float}
\usepackage{booktabs}
\usepackage[version=3]{mhchem}

\makeatletter
\def\@email#1#2{%
 \endgroup
 \patchcmd{\titleblock@produce}
  {\frontmatter@RRAPformat}
  {\frontmatter@RRAPformat{\produce@RRAP{*#1\href{mailto:#2}{#2}}}\frontmatter@RRAPformat}
  {}{}
}%
\makeatother

\begin{document}
\preprint{}

\title{STM study of single phosphorus incorporation into silicon by heating PBr$_3$ on Si(100)}

\author{Tatiana V. Pavlova}
 \email{pavlova@kapella.gpi.ru}
\affiliation{Prokhorov General Physics Institute of the Russian Academy of Sciences, Vavilov str. 38, 119991 Moscow, Russia}
\affiliation{HSE University, Myasnitskaya str. 20, 101000 Moscow, Russia}

\author{Vladimir M. Shevlyuga}
\affiliation{Prokhorov General Physics Institute of the Russian Academy of Sciences, Vavilov str. 38, 119991 Moscow, Russia}


\begin{abstract}
The objective of miniaturizing doped areas in silicon, with the ultimate goal of achieving atomic-precision doping, requires a fundamental understanding of the dopant incorporation process at the atomic level. We present a combined scanning tunneling microscopy (STM) and density functional theory (DFT) investigation of single phosphorus atom incorporation into the Si(100) surface. Phosphorus was supplied via PBr$_3$ molecules, which completely dissociate on Si(100) at room temperature. By performing \emph{in situ} annealing within the STM, we directly tracked the same phosphorus atom before and after heating. Upon annealing, the P atom undergoes an exchange with a nearby Si atom, forming a stable P-Si-Br complex with a Br atom located atop the Si atom of the heterodimer. The activation barrier calculated using DFT is consistent with our observation of doping starting at temperatures as low as 175$^{\circ}$C. These results provide detailed atomic-scale insight into the phosphorus incorporation pathway and offer a foundation for improving methods of precise, single-atom doping in silicon.

\end{abstract}

\maketitle

\section{Introduction}

A deeper understanding of phosphorus interaction with silicon surfaces is essential for advancing doping technologies in microelectronics, as phosphorus is the predominant donor dopant in silicon. Typically, phosphine serves as the precursor for silicon doping. The interaction of PH$_3$ with the Si(100) surface has been extensively studied using scanning tunneling microscopy (STM). At room temperature, phosphine partially or fully dissociates \cite{2005Warschkow, 2006Wilson, 2016Warschkow}, and the resulting phosphorus-containing fragments can diffuse across the surface \cite{2016Warschkow}. Upon sample annealing, phosphorus incorporates into the surface as a substitutional impurity  \cite{2003Schofield}. P-Si exchange begins at temperatures around 250$^{\circ}$C \cite{2009McKibbin}. Density functional theory (DFT) calculations support this observation and propose a minimum energy pathway for the P-Si exchange mechanism \cite{2009Bennett}. Following incorporation, phosphorus forms a P-Si heterodimer in the upper silicon layer, while the silicon atoms ejected onto the surface agglomerate to form islands \cite{PhysRevB.69.195303}.

Following the adsorption of PH$_3$ on Si(100) at room temperature, several stable configurations exist, ranging from partially dissociated fragments to the fully dissociated molecule, where the phosphorus atom can occupy various adsorption sites \cite{2016Warschkow}. The most stable configuration is the fully dissociated molecule with the P atom in the end-bridge position. Upon heating, the incorporation mechanism of phosphorus can vary depending on its initial adsorption site, as demonstrated for the specific cases of the end-bridge and bridge positions \cite{2009Bennett}. Previous STM studies \cite{2009McKibbin, PhysRevB.69.195303, 2006Reusch, 2007Reusch} have examined the overall morphology of the Si(100) surface following PH$_3$ adsorption and subsequent annealing, without resolving the specific influence of the initial phosphorus adsorption site.

This work is motivated by the challenge of achieving deterministic, atomic-precision doping in silicon. The creation of atomic-scale devices operating on single impurities requires extremely precise positioning of dopants within the silicon lattice, as a displacement even by a single lattice constant can affect device characteristics \cite{2024Hsueh}. In this context, a detailed, atomic-level understanding of the phosphorus incorporation process becomes critically important. Therefore, the objective of this study was to investigate in STM the incorporation mechanism of a single P atom into the Si(100) surface specifically from its most stable adsorption configuration.

We selected phosphorus tribromide (PBr$_3$) as the phosphorus source, as it is suitable for atomic-precision doping via halogen-based lithography \cite{2024Pavlova, Pavlova2025JCP} and, furthermore, individual bromine atoms are clearly visible in STM at room temperature \cite{2023Shevlyuga}. Our previous studies have identified the surface structures formed following its dissociative adsorption at room temperature \cite{2023Shevlyuga, 2025PavlovaPCCP}. In the present work, we tracked the same individual phosphorus atom before and after its incorporation using STM to elucidate the exchange mechanism. This was achieved by performing \emph{in situ} sample annealing directly within the STM chamber. Combined with DFT calculations, this allowed us to identify the configuration of the inserted phosphorus atom and establish the minimum energy pathway for the P-Si exchange reaction.

\section{Experimental and computational details}

All experiments were conducted within an ultra-high vacuum (UHV) setup with a base pressure of 5$\times$10$^{-11}$\,Torr. The STM measurements were performed using a GPI CRYO system (SigmaScan Ltd.) at an operating temperature of 77 and 300\,K. Boron-doped Si(100) wafers with a resistivity of 1\,$\Omega$\,cm served as substrates. The Si(100) surface was prepared by outgassing the wafer at 870\,K for several days under UHV conditions, followed by a final flash-anneal to 1470\,K. Subsequently, the sample was cooled to room temperature over a period of 30 minutes. Adsorption of PBr$_3$ was performed by exposing the clean Si(100) surface at room temperature to a partial pressure of $3 \cdot 10^{-10}$\,Torr for a duration of two minutes. The sample with adsorbed PBr$_3$ was annealed by direct current heating either inside the preparation or STM chamber. During this procedure inside the STM chamber, the sample position remained unchanged, while the STM tip was retracted several microns away from the surface. To calibrate the relationship between the heating current and the resulting sample temperature, we conducted a series of experiments on a test silicon wafer. The wafer was heated with a defined current while a thermocouple was pressed to its surface, allowing temperature measurement across the range from 150 to 400$^{\circ}$C. The temperature of 400$^{\circ}$C was verified using an optical pyrometer. Residual phosphorus was removed from the sample between experiments by annealing at 1170\,K overnight \cite{2005Brown}. The STM measurements employed mechanically cut Pt-Rh and Pt-Ir tips, as well as electrochemically etched polycrystalline tungsten (W) tips. A bias voltage ($U_s$) was applied to the sample. All STM images were processed using the WSXM software package \cite{WSXM}.

Spin-polarized DFT calculations were performed using the Perdew-Burke-Ernzerhof (PBE) functional \cite{1996Perdew} within the Vienna \textit{ab initio} simulation package (VASP) \cite{1996Kresse, 1999Kresse}. A plane-wave basis set with a kinetic energy cutoff of 350\,eV was employed. The Si(100) surface was modeled using an eight-layer slab in a 4$\times$4 surface supercell, with a vacuum gap of 14\,{\AA}. A 8$\times$8 supercell was used to model surface structures with a small Si island. Phosphorus and bromine atoms were placed on the top surface, which retained the 2$\times$1 reconstruction, while dangling bonds on the bottom surface were terminated with hydrogen atoms. The atomic positions in the bottom three Si layers were fixed at their bulk values, whereas all other atoms were fully relaxed until the Hellmann-Feynman forces were below 0.01\,eV/\,{\AA}. Brillouin zone integration was performed using a 3$\times$3$\times$1 k-point mesh. STM-images were generated within the Tersoff-Hamann approximation \cite{1985Tersoff}. STM images were simulated for two different buckling configurations of Si dimers and then averaged for comparison with the experiment. Activation energy barriers for atomic transitions were calculated using the nudged elastic band (NEB) method \cite{1998NEB}. For these calculations, the force convergence criterion was set to 0.03\,eV/\,{\AA}. The reaction pathway was calculated using five intermediate images between the initial and final states.

\section{Results and Discussion}

\subsection{Surface annealing experiments}

\begin{figure*}[t!]
\begin{center}
\includegraphics[width=0.8\linewidth]{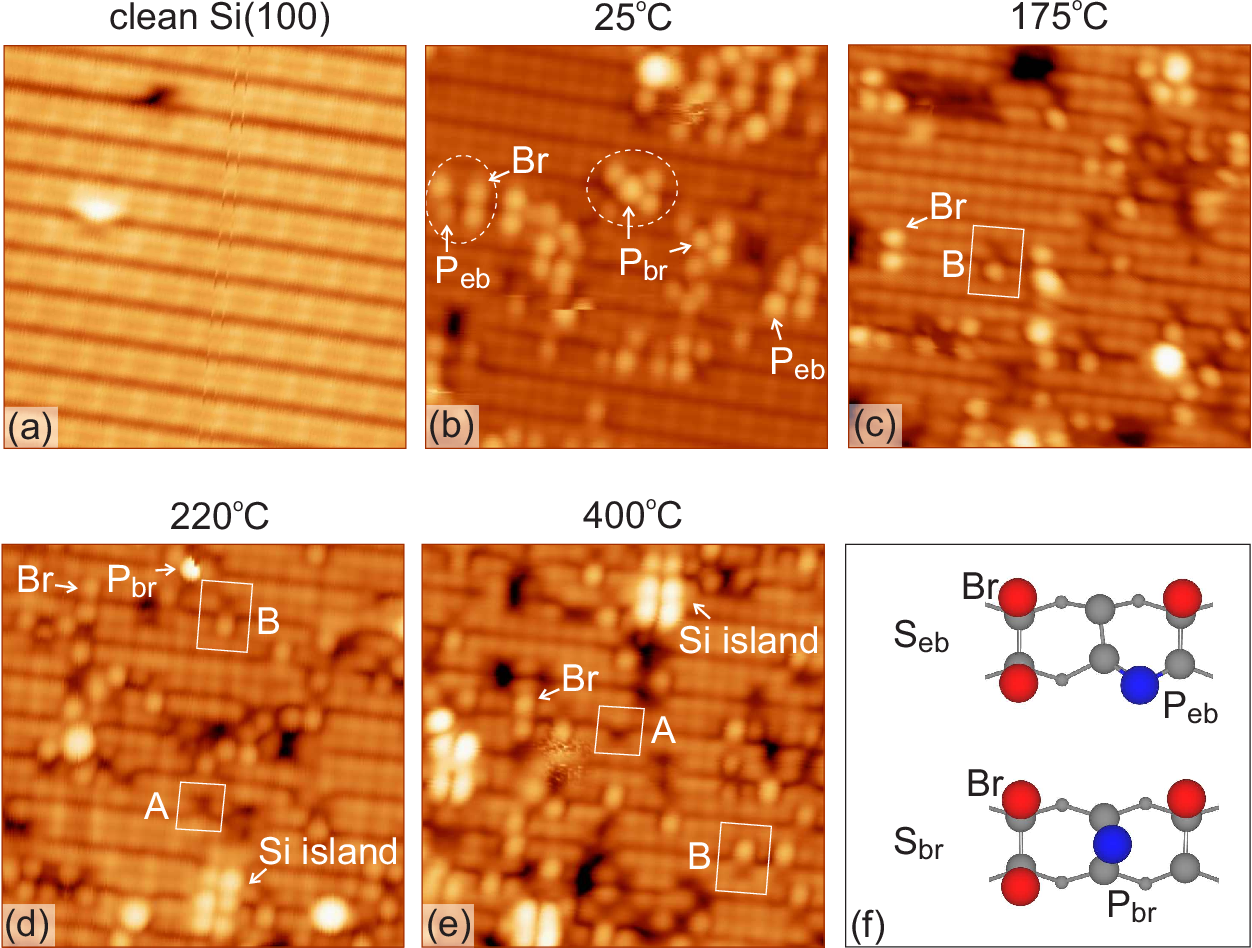}
\caption{\label{fig1}
Adsorption of PBr$_3$ on Si(100)(a) at room temperature (b) followed by consecutive 5-minute anneals to 175$^{\circ}$C (c), 220$^{\circ}$C (d), and 400$^{\circ}$C (e). Empty-state STM images (7.5$\times$7.5\,nm$^2$) were acquired at $U_s = +2$\,V, $I_t$ = 2\,nA, and a sample temperature of 77\,K. Following dissociative adsorption of PBr$_3$, P atoms are observed in the end-bridge ($P_{eb}$) and bridge ($P_{br}$) positions. Br atoms occupy on-top sites above Si atoms, predominantly in pairs on a single Si dimer. Annealing leads to the formation of silicon islands. Depressions, which are suggested to be embedded P atoms, are marked with squares (object A). Depressions adjacent to a protrusion, are marked by rectangles (object B). (f) Top-view models of $S_{eb}$ and $S_{br}$  surface structures formed after  PBr$_3$ dissociation on Si(100). Si atoms are marked in gray for the top layer (large circles) and the next layer (small circles), Br in red, and P in blue. }
\end{center}
\end{figure*}

To determine the temperature at which phosphorus incorporates into the silicon surface, we adsorbed PBr$_3$ on Si(100) at room temperature and sequentially annealed the sample to 175, 220, and 400$^{\circ}$C in the preparation chamber, each annealing step lasting 5 minutes (Fig.~\ref{fig1}). A small number of typical defects are present on the clean surface (Fig.~\ref{fig1}a) .Following PBr$_3$ adsorption on Si(100) at room temperature, the molecule dissociates into individual atoms (Fig.~\ref{fig1}b) \cite{2023Shevlyuga}. Phosphorus and bromine atoms can be easily distinguished in STM images due to their occupation of different adsorption sites on the Si(100) surface. The most stable position for phosphorus is the end-bridge site within the groove between the dimer rows, where it bonds to two Si atoms from adjacent dimers in the same row and to a Si atom from the underlying layer (P$_{eb}$ in Fig.~\ref{fig1}f) \cite{2023Shevlyuga}. This configuration is further stabilized by a Br atom terminating a Si atom of the adjacent dimer  \cite{2025PavlovaPCCP}. The next most favorable position of phosphorus is at the center of a silicon dimer (P$_{br}$ in Fig.~\ref{fig1}f) \cite{2025PavlovaPCCP}. Note that the brightness of phosphorus in the same adsorption positions can be different (Fig.~\ref{fig1}b). Bromine atoms are adsorbed on the Si(100) surface on top of Si atoms.

After sample annealing from 175 to 400$^{\circ}$C (Figs.~\ref{fig1}c--e), Br atoms remained on the surface. This is consistent with the fact that bromine at sub-monolayer coverage begins to desorb slowly from the Si(100) surface at temperatures around 525$^{\circ}$C \cite{PhysRevB.49.11071}. We readily identify Br atoms in the post-anneal STM images because they appear identical to Br atoms prior to heating, occupying on-top positions above Si atoms and frequently observed as pairs on a single dimer. This observation agrees with the established understanding that the most stable adsorption configuration for a bromine pair is atop the atoms of a single Si dimer \cite{2004Lee}. Embedded phosphorus atoms are reported to appear as depressions \cite{2007Reusch, 2006Radny}, and we identified such sites after annealing (object A, marked by a square in Figs.~\ref{fig1}d, e), often in proximity to a protrusion (object B, marked by a rectangle in  Figs.~\ref{fig1}c--e). Note that all STM images were acquired in empty states; therefore, the protrusions and depressions shown in the article correspond only to this polarity. At annealing temperatures of 175$^{\circ}$C and 220$^{\circ}$C, we observed P atoms that had not yet incorporated into the surface (Fig.~\ref{fig1}d).

Silicon islands first emerge at 175$^{\circ}$C, though their density remains extremely low. Following annealing to 220$^{\circ}$C, the island concentration increases (Fig.~\ref{fig1}d), and at 400$^{\circ}$C both the number and size of the islands grow substantially (Fig.~\ref{fig1}e). Previous work \cite{2009McKibbin} also reported the observation of Si islands on Si(100) upon heating the sample with adsorbed PH$_3$  starting from 250$^{\circ}$C. The presence of Si islands is considered an indicator of phosphorus incorporation since P atoms eject Si atoms from the upper layer, occupying their lattice sites \cite{2003Schofield}. Note that the islands formation depends on the concentration of ejected Si atoms from the top layer and also on their mobility.

In this study, our aim was to directly monitor P incorporation and unambiguously identify the incorporation sites, rather than relying on the indirect evidence of island formation. However, from the STM images in Figs.~\ref{fig1}c--e is difficult to identify the P incorporation sites due to the presence of numerous surface defects. Nevertheless, from sequential annealing experiments of the surface with adsorbed PBr$_3$ (Fig.~\ref{fig1}), the appearance of Si islands allowed us to determine that the earliest phosphorus incorporation events can occur at 175$^{\circ}$C. To identify the sites of P incorporation into the surface, we performed an \emph{in situ} STM experiment.

\begin{figure*}[t!]
\begin{center}
\includegraphics[width=\linewidth]{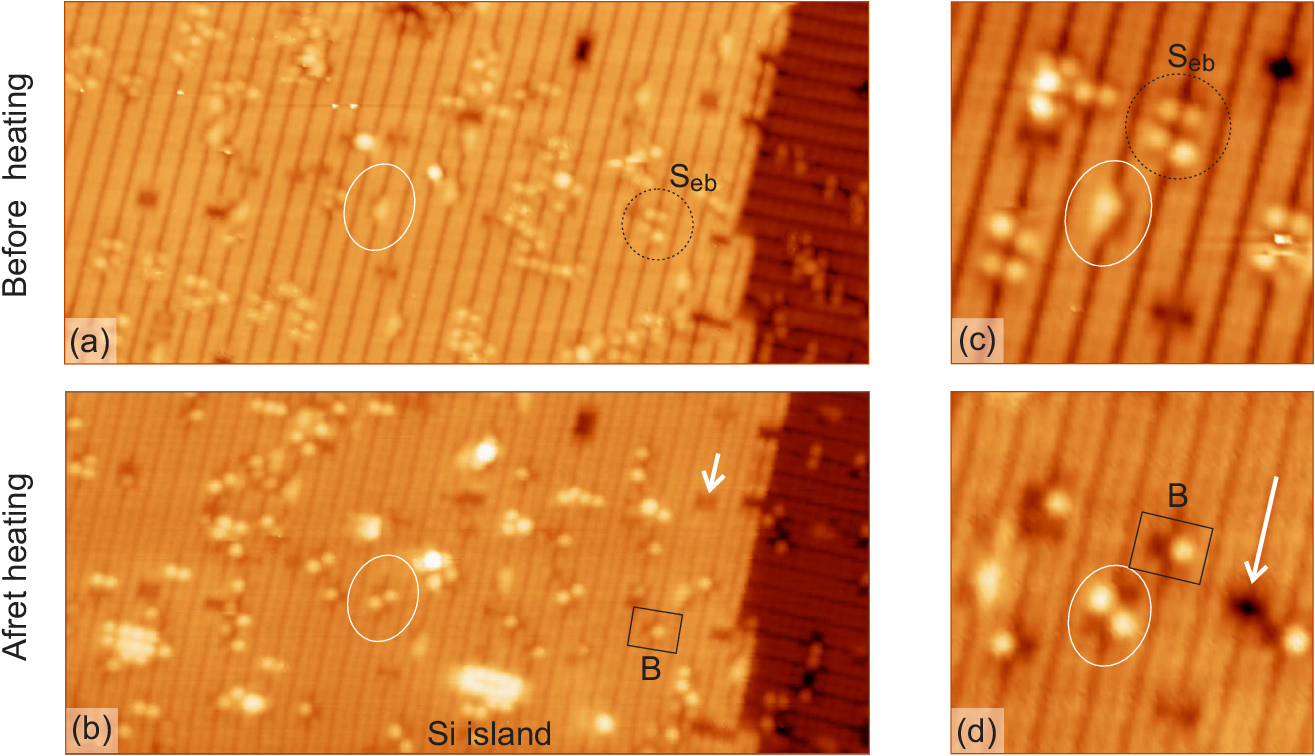}
\caption{\label{fig2}
(a) The Si(100) surface after PBr$_3$ adsorption. (b) The same area following annealing at 300$^{\circ}$C for 1 minute, revealing phosphorus incorporation. Frames (c, d) show atomic-resolution STM images before and after heating. In (a), structure S$_{eb}$ is indicated by a dotted line. In (b) and (d), a depressions adjacent to a protrusion is marked by a rectangle (object B), and a white arrows show the displacement of a Si bivacancy after heating. The white contour indicates water fragments OH and H before and after annealing. Empty-state STM images, 27$\times$12\,nm$^2$ (a, b) and 6.2$\times$6.2\,nm$^2$ (c, d), were acquired at $U_s =+2$\,V, I$_t$ = 2\,nA, and a temperature of 300\,K.
}
\end{center}
\end{figure*}

To study the exchange of phosphorus with silicon following PBr$_3$ molecule dissociation, we recorded STM images of the same surface area before (Fig.~\ref{fig2}a) and after a one-minute anneal at 300$^{\circ}$C inside the STM chamber (Fig.~\ref{fig2}b). To find the same area after heating, we used atomic steps as reference points, as their shape remained unchanged. Following the anneal, numerous changes were observed on the surface, in particular, silicon islands appeared (Fig.~\ref{fig2}b). At the locations of dissociated PBr$_3$ molecules in structure S$_{eb}$ (black circle in Fig.~\ref{fig2}a), features consisting of a depression and a protrusion on a single Si dimer appeared in most cases (object B, marked by a rectangle in Fig.~\ref{fig2}b). At the locations of dissociated water molecules, two depressions and two protrusions, arranged diagonally, appeared. The displacement of Si bivacancy after heating was also observed (indicated by a white arrow in Fig.~\ref{fig2}d ). Figures~\ref{fig2}c, d shows atomic-resolution STM images of dissociated PBr$_3$ and H$_2$O molecules in the same surface area before and after heating. We would like to emphasis that, according to our experimental data, object B was observed at the same place where the PBr$_3$ molecule had originally adsorbed. More specifically, object B was found at the location of the Si dimer on both sides of the end-bridge position of the P atom.

\subsection{Identification of inserted phosphorus}

\begin{figure*}[t!]
\begin{center}
\includegraphics[width=\linewidth]{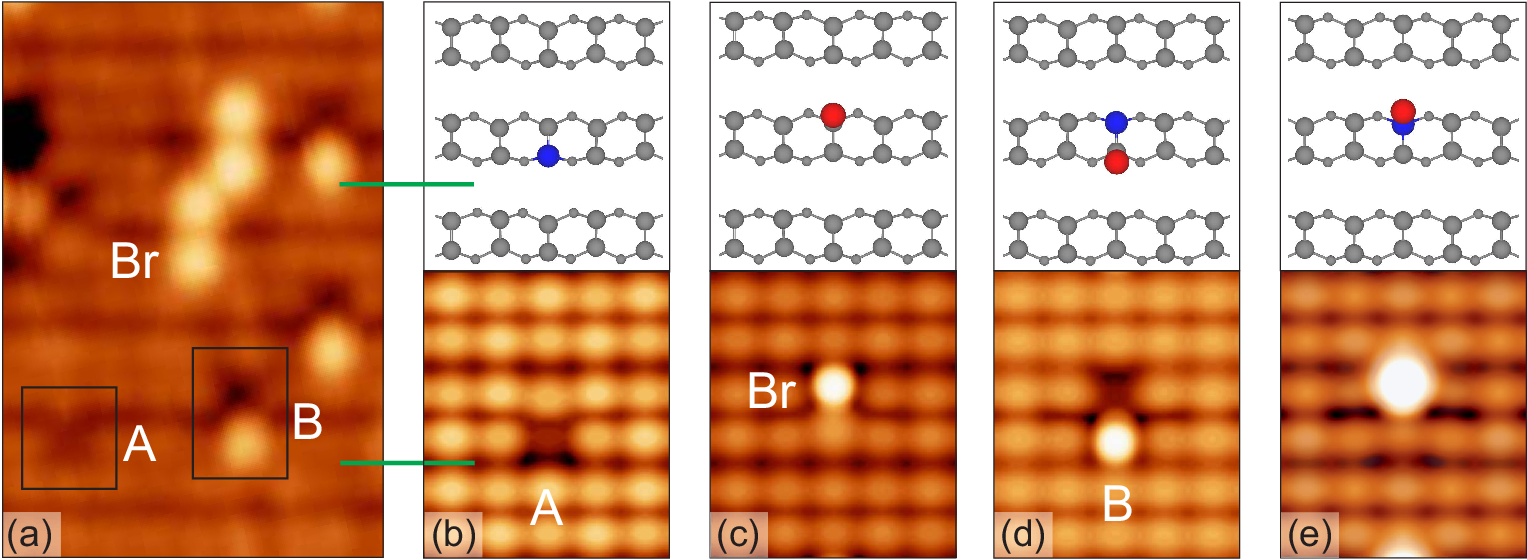}
\caption{\label{fig3}Identification of the phosphorus incorporation sites. (a) Empty-state STM image ($U_s =+2.5$\,V, I$_t$ = 3.0\,nA, 300\,K) of the Si(100) surface with adsorbed PBr$_3$ after annealing to 300$^{\circ}$C for 1 minute. (b--e) Top-view models and simulated empty-state STM images with an embedded P atom and/or an adsorbed Br atom. Si atoms are marked in gray for the top layer (large circles) and the next layer (small circles), Br in red, and P in blue. The green lines in (a) and (b) separate the dimer rows.
 }
\end{center}
\end{figure*}

To interpret objects A and B observed after heating a dissociated PBr$_3$ molecule (Fig.~\ref{fig3}a), we performed calculations of various configurations with a P atom substituting a Si atom. An embedded P atom in the Si(100) surface appears as a depression in the simulated STM image (Fig.~\ref{fig3}b), resembling object A in the experimental STM image (Fig.~\ref{fig3}a). An adsorbed Br atom appears as a protrusion (Fig.~\ref{fig3}c), consistent with the objects we identify as Br atoms in the experimental STM image (Fig.~\ref{fig3}a). If a Br atom is placed atop a heterodimer (P-Si dimer), the embedded P atom appears darker than the Si surface atoms, while the Br atom appears bright (Fig.~\ref{fig3}d), matching the experimental STM image of object B (Fig.~\ref{fig3}a).

The heterodimer with a Br atom atop the Si atom (P-Si-Br) turned out to be the most stable structure among the calculated configurations involving embedded phosphorus and adsorbed bromine. In particular, the structure with a Br atom on the P atom of the heterodimer (Fig.~\ref{fig3}e) was found to be less favorable by 1.26 eV. If the embedded P and the Br atom are positioned far from each other, such a structure is less stable than the P-Si-Br structure by 0.35 eV, indicating that it is energetically more favorable for the Br atom to locate atop the Si atom of the heterodimer than at a distance from it. Consequently, upon annealing the dissociated PBr$_3$ molecule, the P-Si-Br structure forms, as it is the most stable configuration and the experimental (object B in Fig.~\ref{fig3}a) and simulated (Fig.~\ref{fig3}d) STM images are in agreement. Heterodimers without Br atoms were also observed (object A in Fig.~\ref{fig3}a), but less frequently, in line with this structure being less energetically favorable than the heterodimer with a Br atom.

Note that the formation of P-Si heterodimers has been reported in a number of studies \cite{PhysRevB.69.195303, 2007Reusch, 2006Radny}, where PH$_3$ was used as the phosphorus source. In Ref. \cite{2006Radny}, the heterodimer was also visualized in empty-state STM images at voltages > 2 V as a depression, similar to our case (object A in Figs.~\ref{fig3}a,b). However, at voltages of +1.3 V and below, a protrusion appeared adjacent to the depression, which was interpreted as the charge on the silicon dangling bond of the heterodimer \cite{2006Radny}. This is a different situation, as we observe the protrusion at voltages > 2 V, and it originates from a Br atom on the Si atom of the heterodimer (object B in Figs.~\ref{fig3}a,d).

\begin{figure}[h]
\begin{center}
\includegraphics[width=\linewidth]{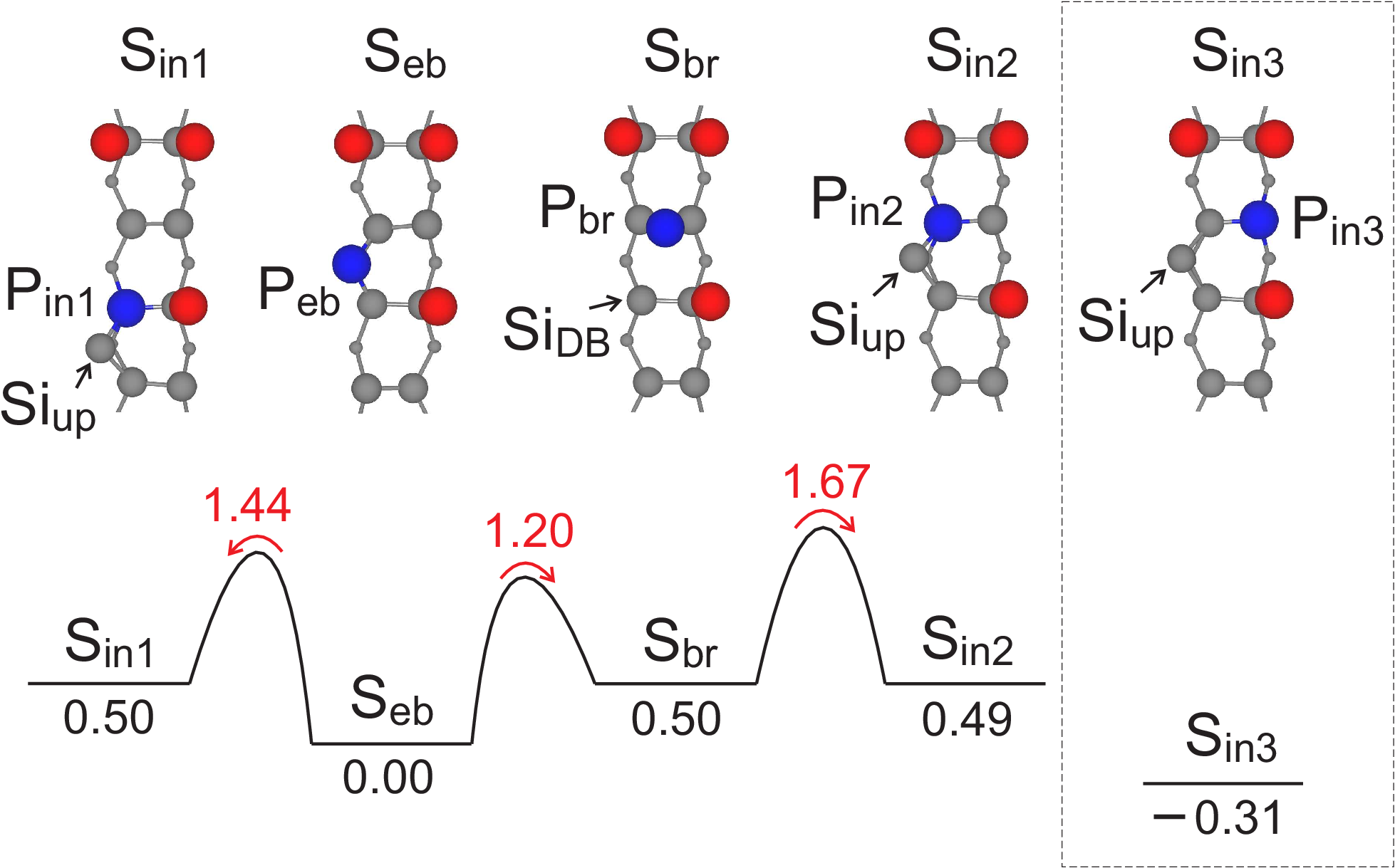}
\caption{\label{fig4}Energy barrier diagram for the P incorporation upon adsorption of the PBr$_3$ molecule on  Si(100) from end-bridge P$_{eb}$ and bridge P$_{br}$ positions. The most favorable positions for phosphorus adsorption (S$_{eb}$ and S$_{br}$) and incorporation (S$_{in1}$,  S$_{in2}$, and S$_{in3}$) are shown on top. Although S$_{in3}$ is the most energetically favorable configuration, the reaction pathway leading to it was not calculated, as it was not observed experimentally. Energies are given in electronvolts. The red number denotes the activation barrier, values are given relative to S$_{eb}$ . Si atoms are marked in gray for the top layer (large circles) and the next layer (small circles), Br in red, and the P in blue.
}
\end{center}
\end{figure}

Having identified the structures with a P atom before and after heating, we calculated the activation barrier for the P incorporation in the presence of three Br atoms (Fig.~\ref{fig4}). In the initial structure S$_{eb}$, the phosphorus is in the end-bridge position (P$_{eb}$ ), while in the final structure S$_{in1}$, the phosphorus is in a heterodimer(P$_{in1}$) with a Br atom on the Si atom. Upon incorporation, the P atom substitutes for a Si atom in the dimer, with the displaced Si atom occupying the end-bridge position (Si$_{up}$ ). When phosphorus displaces a Si atom from a dimer that is terminated at the opposite end by a Br atom, the activation barrier is 1.44 eV. We also calculated the P-Si exchange pathway from the end-bridge (P$_{eb}$ ) and bridge (P$_{br}$) positions by analogy with the most favorable pathway in a similar structure according to Ref.~\cite{Pavlova2025JCP}. From the end-bridge position, phosphorus diffuses to the bridge, and then displaces one of the underlying Si atoms. The phosphorus preferentially displaces the Si atom adjacent to a Si atom with a dangling bond (Si$_{DB}$ ) \cite{Pavlova2025JCP}. The activation barrier for incorporation from P$_{br}$ into position P$_{in2}$ is 1.17 eV. Taking into account diffusion from P$_{eb}$  to P$_{br}$ , the overall barrier from the end-bridge position to P$_{in2}$ is 1.67 eV. Thus, the minimum energy pathway from P$_{eb}$  leads to P$_{in1}$ configuration with an activation barrier of 1.44 eV.

The final structures S$_{in1}$ and S$_{in2}$ are less stable compared to the initial one (S$_{eb}$). Although we constructed a structure more stable than the initial one (S$_{in3}$ in Fig.~\ref{fig4}), we have not identified such a structure in the experimental STM images. Accordingly, we modeled a structure after P incorporation that accounts for silicon island formation, given that Si atoms are experimentally observed to combine into islands. For this, we increased the supercell size by a factor of 4 and placed four S$_{eb}$ structures. In the final state, we placed four S$_{in1}$ structures, from which we formed an island of four ejected silicon atoms Si$_{up}$. The final state with embedded phosphorus in the P-Si-Br heterodimer and a Si island is more stable than the initial S$_{eb}$ state by 0.41 eV per one dissociated PBr$_3$ molecule. Thus, phosphorus incorporation accompanied by the formation of Si islands results in a more stable surface structure, in agreement with experimental observations.

The process of P-Si exchange has been studied in other works, with and without an adsorbate \cite{2016Warschkow, 2009Bennett, Pavlova2025JCP}. In particular,  the activation barrier for P-Si exchange was calculated on a clean Si(100) surface starting from the bridge position, yielding a barrier of 1.24 eV, and from the end-bridge position, yielding a barrier of 1.55 eV \cite{2009Bennett}. The estimated temperatures for phosphorus incorporation were 420 K and 530 K, respectively. Note that the work \cite{2009McKibbin} reported island formation on Si(100) with adsorbed phosphine after heating starting from 250$^{\circ}$C, which also aligns with the temperature estimates provided in the calculations of Ref.~\cite{2009Bennett}. Consequently, the barriers obtained in our work for P-Si exchange from the bridge position, 1.17 eV, and from the end-bridge position, 1.44 eV, are consistent with the fact that we began to observe heterodimers at temperatures of 175$^{\circ}$C.

\subsection{Formation of phosphorus dimers}

\begin{figure}[h]
\begin{center}
\includegraphics[width=\linewidth]{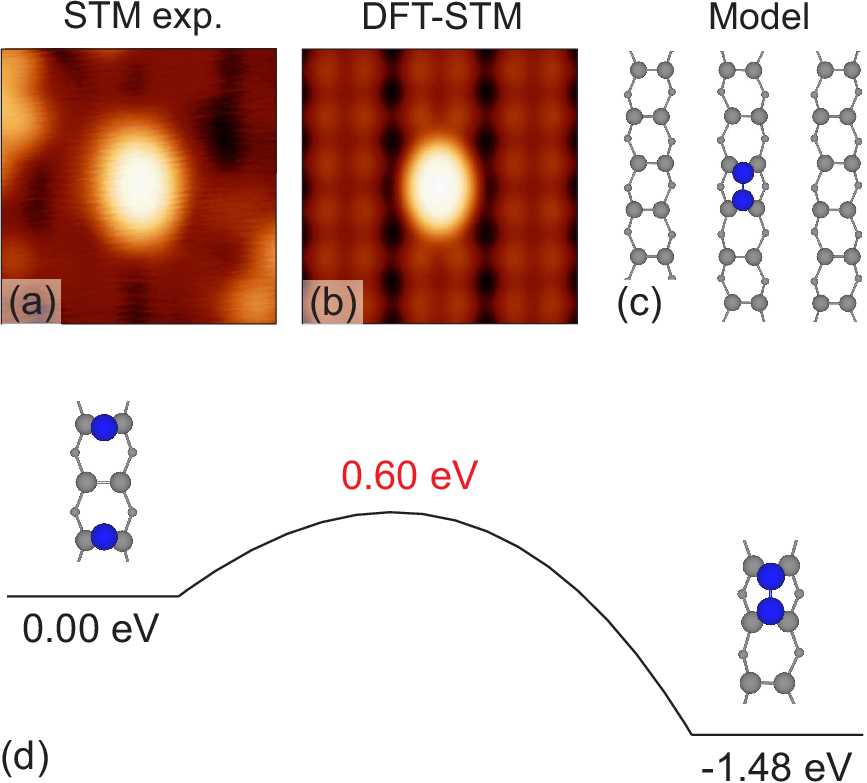}
\caption{\label{fig5} Phosphorus dimer on the Si(100) surface. (a) Empty-state STM image ($U_s =+3.2$\,V, I$_t$ = 3.0\,nA, 77\,K) of one of the bright objects formed after  PBr$_3$ adsorption on Si(100) at room temperature. (b) Simulated empty-state STM image and (c) top-view model of the most stable configuration of P$_2$ on Si(100). (d) Energy barrier diagram for the P association from two P atoms in bridge positions within the dimer row. The red number denotes the activation barrier. Si atoms are marked in gray, Br atoms in red, and the P atom in blue.}
\end{center}
\end{figure}

STM images of the Si(100) surface following PBr$_3$ adsorption and subsequent annealing show bright objects elongated along the dimer rows (Fig.~\ref{fig5}a). We attribute some of these bright objects to phosphorus dimers (P$_2$) on the Si(100) surface. Figure~\ref{fig5}b shows a simulated STM image of such a dimer, which is consistent with the STM image of the bright object (Fig.~\ref{fig5}a). In the most favorable position of P$_2$ on Si(100), it is aligned along a dimer row, with its center located between dimers within the same row (Fig.~\ref{fig5}c) \cite{2013Sagisaka}. According to our calculations, the formation of phosphorus dimers is energetically more favorable than two individual P atoms in bridge positions separated by one dimer by 1.48\,eV (Fig.~\ref{fig5}d). This is consistent with previous work, which also reported an energy gain for the pairing of two P atoms from end-bridge positions \cite{2010Bennett}. We note that, based on our calculations, the formation of trimers (P$_3$) and tetramers (P$_4$) is less favorable than dimer formation, in agreement with Ref.~\cite{2013Sagisaka}. The activation energy for the association of two P atoms into a dimer is 0.60 eV. This energy is comparable to the activation energy for phosphorus diffusion within a dimer row, 0.79 eV \cite{2025PavlovaPCCP}. However, only P atoms located within the same dimer row without bromine are able to pair up since Br atoms substantially reduce the mobility of phosphorus \cite{2025PavlovaPCCP}. Therefore, this suggest that after PBr$_3$ adsorption on Si(100), some of the P atoms can combine into dimers. Another interpretation of the bright objects could be silicon addimers, which are more stable than two single silicon atoms and occupy the same position as phosphorus dimers. Nonetheless, although Si$_2$ may form on the surface after annealing, following phosphorus incorporation into the silicon substrate, they are not expected to appear prior to annealing of the PBr$_3$-covered surface.

We estimated the possibility of phosphorus incorporation into silicon from a P$_2$ dimer. The energy required to dissociate the dimer into two P$_{br}$ atoms in bridge positions is 2.08 eV (the reverse reaction in Fig.~\ref{fig5}d). Following this, P$_{br}$ can incorporate into the upper silicon layer at the P$_{in2}$ site with a barrier of 1.17 eV (Fig.~\ref{fig4}). Another pathway, involving dissociation of the P$_2$ dimer into two P$_{eb}$ located one dimer apart within the same dimer row, can be estimated from a previous theoretical study \cite{2010Bennett}, which reported an activation barrier of 1.82 eV for this process. Subsequently, P$_{eb}$ can incorporate into the upper silicon layer at the P$_{in1}$ site with a barrier of 1.44 eV (Fig.~\ref{fig4}). In both cases, the energy required for dimer dissociation exceeds the activation energy for phosphorus incorporation into silicon. Thus, if we take into account the possibility of phosphorus to form dimers, higher temperatures may be required for their subsequent incorporation. This is due to an additional activation barrier for dimer dissociation, which exceeds the incorporation barrier for isolated phosphorus atoms.

\subsection{A pair of bromine atoms}

A direct comparison of the same surface area before and after annealing demonstrates that phosphorus mostly incorporates at its original adsorption site, while the bromine atom pairs are no longer present in the vicinity. Indeed, from the initial S$_{eb}$ structure, consisting of a P atom and three Br atoms (Fig.~\ref{fig2}c), the P atom became incorporated, with one bromine atom remaining adjacent to it (Fig.~\ref{fig2}d). The two remaining Br atoms diffused away across the surface, as they are no longer present near the heterodimer. To determine whether phosphorus incorporation occurs prior to bromine diffusion or vice versa, we calculated the activation barrier for the displacement of a Br-Br pair between adjacent dimers. We did not consider the separation of the Br pair into single Br atoms on different dimers, as post-anneal STM images show Br atoms still predominantly paired on a single dimer. If a pair of Br atoms moves simultaneously from one dimer to another, the barrier is 2.14 eV. However, if one Br atom moves to a neighboring dimer first, followed by the second Br atom, the diffusion barrier is 0.93 eV. Thus, two bromine atoms diffuse sequentially. The activation barrier for sequential bromine diffusion is significantly lower than the barrier for P-Si exchange; consequently, bromine can diffuse at a lower temperature than phosphorus incorporates.

After annealing, a dissociated water molecule (OH and H) appears in the STM image as two protrusions and two depressions arranged diagonally (Figs.~\ref{fig2} c,d). Although this diagonal object looks similar to two bromine-terminated heterodimers (Fig.~\ref{fig3} d) positioned adjacent to each other, OH and H were present at this location prior to heating. We tentatively propose, based on preliminary energy estimates, that the mobile Br pair attaches to OH and H during annealing, as the resulting complexes are energetically favorable. Furthermore, a phosphorus atom may diffuse toward an oxygen atom, resulting in a more stable adsorption position of phosphorus \cite{2025PavlovaPCCP}. The observation that the diagonal object is not just two Br-terminated heterodimers was made possible only by performing the annealing inside the STM chamber, followed by imaging the same surface area. It is important to highlight this fact to avoid confusing water-related features with two Br-terminated heterodimers.

\section{Conclusions}

In this study, we employed STM to investigate the incorporation of phosphorus from dissociated PBr$_3$  molecules into the Si(100) surface. By annealing the sample directly inside the STM chamber, we successfully tracked individual phosphorus atoms and identified the resulting structures after incorporation. Phosphorus incorporates by forming a P-Si-Br heterodimer, with a bromine atom on the silicon atom of the heterodimer. DFT calculations yield a minimum activation barrier of 1.44 eV for the P transition from the initial end-bridge adsorption site to the final brominated heterodimer structure observed experimentally. This relatively low barrier is consistent with the onset of incorporation already at 175$^{\circ}$C. Phosphorus atoms may also pair into dimers; however, incorporation from such dimers requires a higher activation energy due to the need for prior dimer dissociation. Upon annealing, complexes containing dissociated water and possibly Br and P atoms can form at the original water adsorption sites. Although these complexes resemble two diagonally arranged brominated heterodimers in STM images, they do not correspond to such structures. These findings enable direct atomic-scale identification of phosphorus incorporation from PBr$_3$ in STM, avoiding the need for indirect indicators such as silicon island formation.

\section{Acknowledgments}
This research was supported in part through computational resources of HPC facilities at HSE University.

\bibliography{ApplSurfSci_P_incorp_rev_arxiv}

\end{document}